\documentclass[a4paper,11pt]{article}
\usepackage{arxiv}

\usepackage{amsmath,amsfonts,amssymb}
\usepackage{graphicx}
\usepackage[colorlinks=true, allcolors=blue]{hyperref}
\usepackage{threeparttable} 
\usepackage{macro_ref}
\usepackage{geometry}
\title{Silencing the stellar glare with kernel-nulling and active phase control on photonic chips}
\renewcommand{\shorttitle}{Silencing the stellar glare with kernel-nulling and active phase control on photonic chips}
\date{}

\begin{document} 
\maketitle

\vspace{-6em}
\begin{center}
Marc-Antoine Martinod$^a$
Vincent Foriel$^a$
Nick Cvetojevic$^a$
Frantz Martinache$^a$
Jeronimo Calderon-Gomez$^a$
Roxanne Ligi$^a$
David Mary$^a$

~\newline

$^a$Université Côte d'Azur, Observatoire de la Côte d'Azur, CNRS, Laboratoire Lagrange, France
\end{center}
~\vspace{-0.5em}

\begin{abstract}
Characterizing exoplanets and finding biosignatures are among the hottest topics in astronomy. 
This quest is challenged by the brightness contrast between the faint planet and the overwhelming glare of the host star, their small angular separation, and observing conditions degraded by optical aberrations. 
Nulling interferometry is a promising technique to achieve this goal by fulfilling requirements for both high contrast and high angular resolution. 
Photonic Integrated Circuits (PICs) provide compactness, design versatility, and scalability needed for nulling. 
With the PHOTONICS project, we aim to create novel coherent combination architectures, based on a multimode interferometer (MMI) coupler and on-chip active phase control with no moving parts. 
We present the first lab tests of a four-beam kernel-nuller made out of a single MMI coupler with active phase control: ABCD fringe-tracking and the characterization of the contrast detection limits of a four-beam kernel-nuller.
\end{abstract}

\section{Introduction}
\label{sec:intro}
Integrated optics and photonic technologies are becoming more widely used in astrophysics, particularly in high angular resolution and high contrast imaging.
Photonics-based solutions are smaller, lighter and cheaper than their bulk-optics equivalents, while simultaneously delivering functionality that is otherwise difficult or even impossible to achieve (e.g., spatial filtering).
Once starlight is injected into photonic circuits, subsequent losses can be low and complex sequences of optical processing can be performed with no possibility for misalignment or drift. 
Thus, significant improvements on the instrument stability are possible.
Unprecedented precision has been reached by deploying such components in interferometry in the visible \cite{martinod2018, huby2012} and in the infrared domain \cite{gravity2017, 2019A&A...623L..11G}.
    
The detection and the characterization of exoplanets close to their host star, and particularly within the habitable zone, are among the most pressing instrumental challenges faced by contemporary astronomy.
Such detections require a high angular resolution and the ability to handle a high planet-to-star contrast ratio, the latter ranging from $10^{-4}$, for self-luminous hot exoplanets observed in the mid-infrared \cite{Marois2008}, to $10^{-10}$, for Earth-like exoplanets imaged in reflected light from their host star \cite{Schworer2015}.
Nulling interferometry fulfills both requirements by making an on-axis source (the star) destructively interfere while transmitting the light from an off-axis companion (a planet) \cite{Bracewell1978}.
The huge potential of integrated optics to process the light for this observing technique has been demonstrated by the Guided-Light Interferometric Nulling Technology (GLINT) instrument \cite{lagadec2018, norris2020, martinod2021, Spalding2024, Rossini2026} and it will be the core component of the next generation of exoplanet hunters such as NOTT\cite{hi5_defrere, nott_sanny2026_perf_chip} or Seidr\cite{dahl_seidr2026}.

The ``photonic'' bench at Observatoire de la Côte d'Azur explores the potential of the multimode interferometer (MMI) to perform nulling instead of using evanescent couplers\cite{Chingaipe2022, Cvetojevic2022}.
The MMI enables greater flexibility in its use compared to directional couplers, as a single MMI behaves differently depending on the number of beams feeding it.
A critical use is the application of the concept of kernel-nulling\cite{Martinache2018}, a generalization of the closure phase.
The bench also explores the use of on-chip thermo-optic phase shifters (TOPS) utilizing the thermo-optical effect to tune the differential phase for potential active or adaptive phase controls.
This paper presents the first results of the characterization of the chromatic behavior of the TOPS in phase control and the contrast limit of a 4T nuller made of an MMI.

\section{Nulling and kernel nulling}
\label{sec:theory}

\subsection{Principles of Nulling Interferometry}
Direct imaging of exoplanetary systems is fundamentally limited by the coherent glare of the host star.
Nulling interferometry serves as the interferometric equivalent of coronagraphy, acting as a spatial angular filter.
By introducing a precise $\pi$ phase shift between the optical paths coming from symmetric sub-apertures, the light originating from an on-axis source is canceled out via destructive interference.
Conversely, an off-axis source, such as an orbiting planet, experiences a different geometric path delay which shifts its relative phase away from the destructive fringe.
As a result, the off-axis planetary signal is transmitted to the detection system.

The observable used to quantify the extinction of the light is called the null depth.
This observable relates the residual intensity at the destructive output to the maximum intensity at the constructive output.
This relationship is fundamentally governed by the visibility ($V$) of the fringe pattern, expressed as:
\begin{equation}
    N = \frac{I_{\text{destructive}}}{I_{\text{constructive}}} = \frac{1-V}{1+V}
\end{equation}
With an error-free instrument observing a point source, the visibility equals unity, yielding a perfect null.
In real observing conditions, spatial and temporal wavefront perturbations degrade the coherence of the beams, causing the null depth to leak.

\subsection{Simple and double null depth}
\begin{figure}
    \centering
    \includegraphics[width=0.4\linewidth]{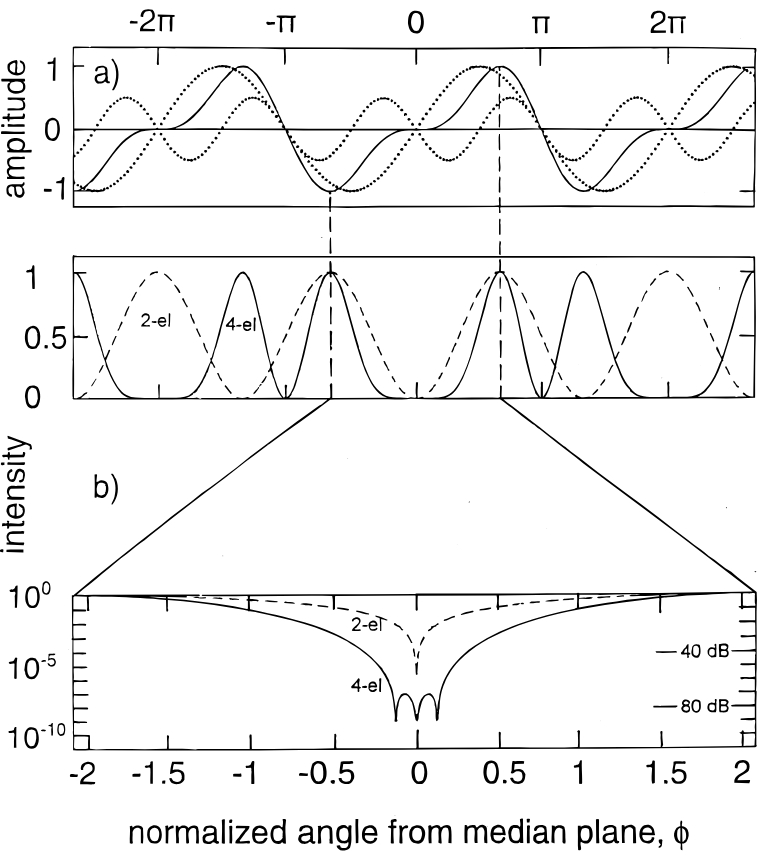}
    \caption{From Angel \& Woolf (1997)\cite{Angel1997} Extinction curves with a single Bracewell combiner with two telescopes (dotted line) or a double Bracewell one with four telescopes (solid line). The latter exhibit deeper and wider null depth on sky than the former.}
    \label{fig:angelwoolf}
\end{figure}

Staged nuller has always been expected to deliver better contrast than a single nuller, like the Bracewell architecture, and more robust against leakage\cite{Angel1997} (Fig.~\ref{fig:angelwoolf}).
New generation of nuller\cite{hi5_defrere} will use a double-Bracewell architecture using two stages of nullers.

The alternative presented here is the realization of the theoretical concept of Martinache \& Ireland\cite{Martinache2018} where an MultiMode Interferometer cavity can single-handledly manage to get the double null along with some self-calibrated ``kernel'' observable which is robust against perturbation.
Not only the same observations as for a staged nuller are expected, but also a double null depth should exhibit a chromatic behavior that paves the way to a ``tunable'' nuller where one can choose the wavelength of maximum extinction, with a fine phase control system.

\section{The multimode interferometer (MMI) architecture}
\label{sec:mmi}

Rather than relying on networks of classic directional couplers, this work utilizes a MultiMode Interferometer (MMI) coupler as the combiner.
An MMI consists of a wide, centralized waveguide cavity designed to support a large number of spatial modes, connected to singlemode input and output channels.
When incoming light from the input waveguides enters the multimode section, the discrete wavefront profiles decompose into a linear combination of the cavity's guided eigenmodes.
Because each eigenmode propagates through the cavity at a unique phase velocity governed by its respective propagation constant, they continuously slip in phase relative to one another along the propagation axis.

\begin{figure}[h]
   \centering
   \includegraphics[width=0.3\textwidth]{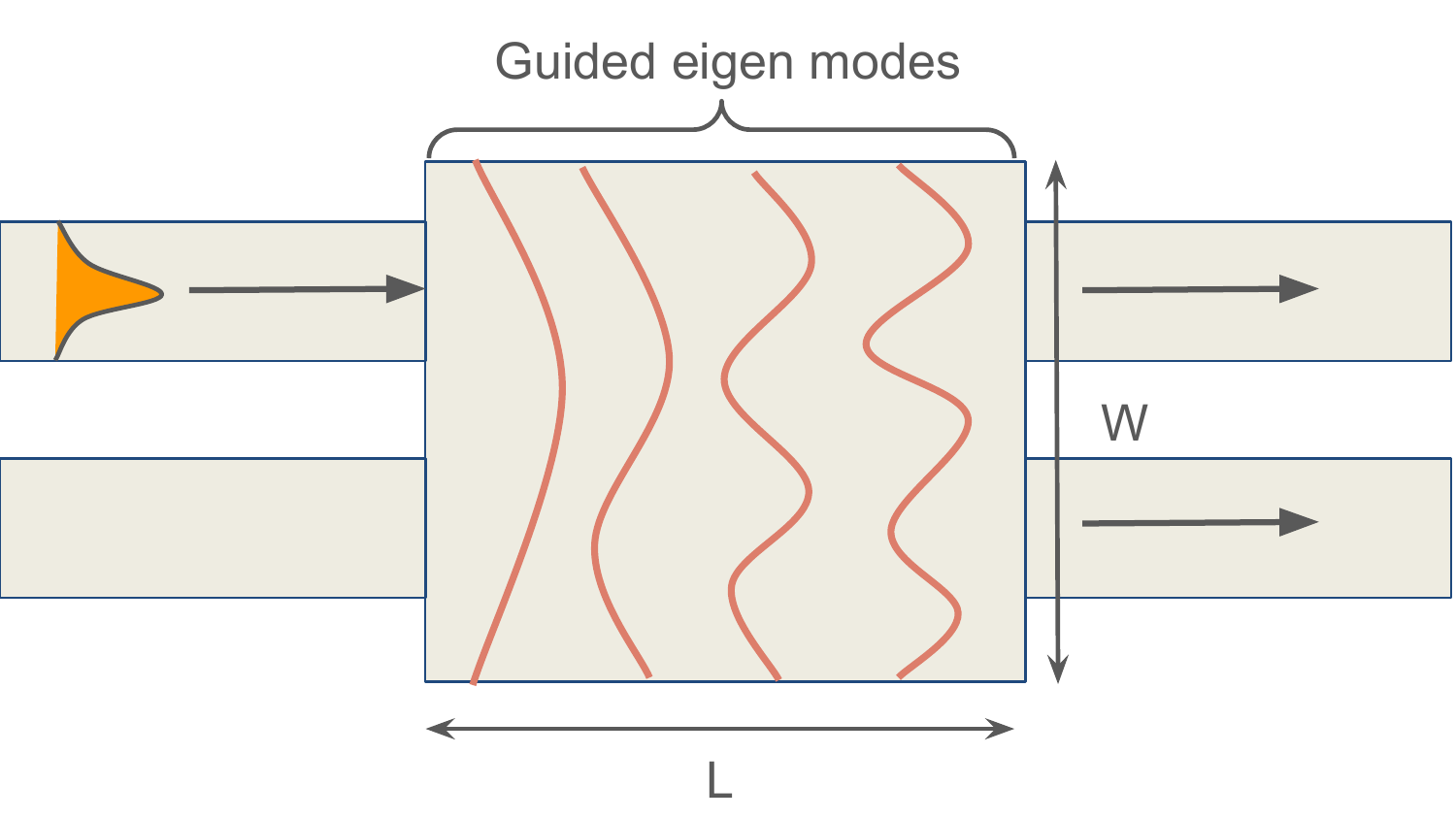}
   \caption{Schematic layout of a MultiMode Interferometer (MMI) cavity. Incoming light splits into multiple guided eigenmodes within the multimode section of width $W$ and length $L$, creating a predictable self-imaging interference pattern at the output plane.}
   \label{fig:mmi}
\end{figure}

As a consequence of this modal dispersion, the modes constructively interfere at highly localized, periodic distances along the length of the cavity.
By precisely engineering the structural width ($W$) and total length ($L$) of the MMI block, a highly symmetric, stable, and multi-port interference pattern is reconstructed.
Singlemode output waveguides are positioned at these precise self-imaging zones to sample and collect the interferometric combinations.
This provides a highly integrated and robust beam-combining architecture on a single monolithic chip.

\section{Description of the Photonics test bench}
\label{sec:bench}

\begin{figure}[h]
    \centering
    \begin{tabular}{c}
        \includegraphics[width=0.55\textwidth]{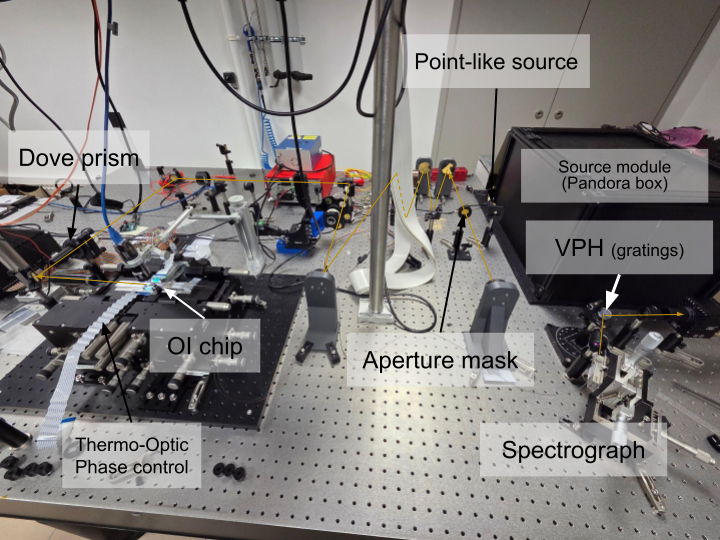}\\
        \includegraphics[width=0.55\textwidth]{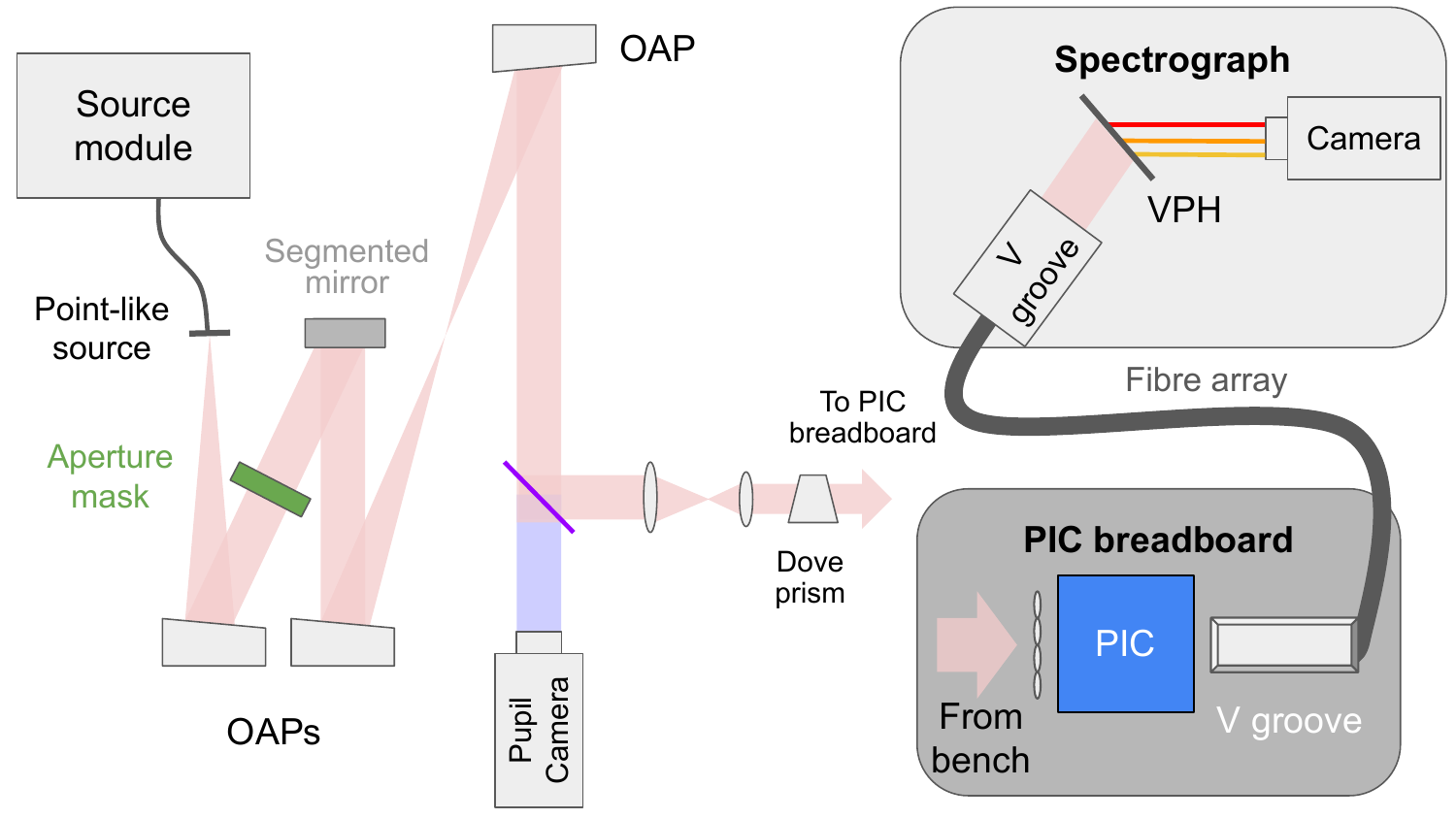}\\
    \end{tabular}
   \caption{Top: Photograph of the experimental PHOTONICS laboratory bench setup.
   Bottom: Simplified diagram of the photonic bench.}
   \label{fig:bench}
\end{figure}

\begin{figure}[h]
    \centering
    \begin{tabular}{cc}
        \includegraphics[width=0.3\linewidth]{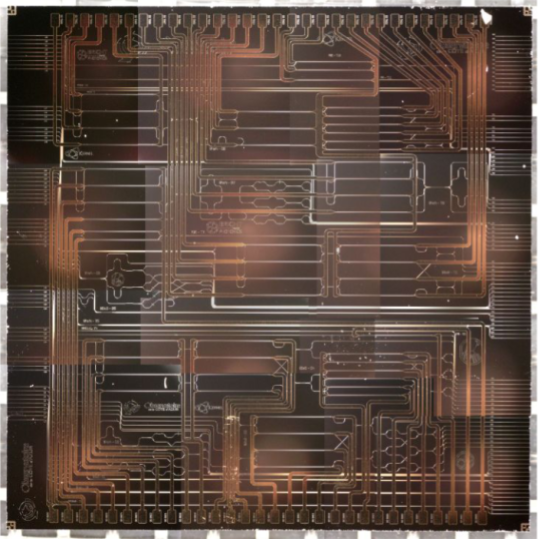}  &  
        \includegraphics[width=0.3\textwidth]{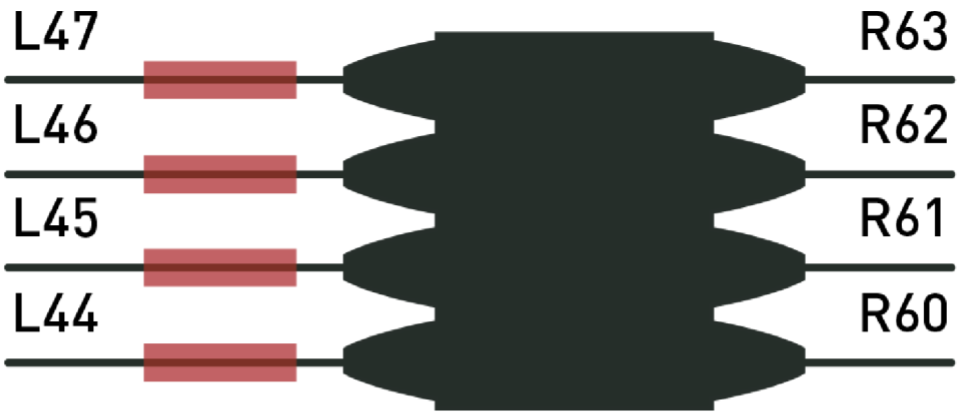}\\
    \end{tabular}
    \caption{Left: picture of the tested chip. Right: diagram of the tested device: a $4\times4$ MMI with thermo-resistance enabling active phase control.}
    \label{fig:chip}
\end{figure}

To validate the performance of the integrated kernel-nuller architecture, a dedicated opto-mechanical characterization bench was assembled\cite{Chingaipe2022}.
This testbed simulates a multi-aperture telescope array and injects the synthesized beams into our custom photonic devices (Fig.~\ref{fig:bench}).
The primary opto-mechanical layout and its key components are outlined below:

\begin{itemize}
    \item \textbf{Source Module:} The bench is fed either by a tunable laser or a super-continuum broadband source, enabling both monochromatic phase sweeps and broad spectral characterization across the H band. The output is a polycrystalline fiber that acts as a point-like source.
    \item \textbf{Aperture Mask:} The light emitted from the source illuminates a Boston Micromachine segmented mirror (also called ``deformable mirror'' or ``DM'') combined with a custom 4-aperture mask.
    This setup splits the continuous wavefront into four discrete coherent beams, mimicking the pupil configuration of an interferometric array.
    The size and spacing of the apertures match those of the mirror segments downstream.
    \item \textbf{Segmented mirror:} These mirrors enable the individual control of piston, tip and tilt of each beam for injection and phase configuration.
    A visible pupil camera images the alignment of the mirror with the mask.
    A Dove prism is integrated into the path to turn the vertical layout of the beams into a horizontal orientation to match the integrated optics bus.
    \item \textbf{Photonic Integrated Circuit (PIC) Chip:} It integrates several optical features onto a single chip (Fig.~\ref{fig:chip}, left) to test various configurations of combiners and on-chip phase control.
    The main device under test is an optical chip featuring a $4\times4$ MMI (Fig.~\ref{fig:chip}, right) designed to coherently mix the four collected sub-apertures with on-chip Thermo-Optical Phase Shifters (TOPS) to finely tune the phase before coherent combination.
\item \textbf{Active Phase Control Hardware:} This includes the driver and electronics required to operate the TOPS.
    These actuators function by locally heating the waveguide channels.
    The localized temperature change alters the material's refractive index via the thermo-optic effect, adjusting the propagation constant and introducing a controlled phase shift\cite{Parra2024}.
    These solid-state actuators require no moving mechanical parts and operate at fast speeds, with a modulation bandwidth reaching approximately $1\text{~kHz}$ according to specifications and 100~Hz in practice due to sub-optimized driver and electronics.
    \item \textbf{Spectrograph:} The outputs of the photonic chip are routed to a spectrograph equipped with a Volume Phase Holographic (VPH) gratings, which disperse the interference pattern across a 180~nm bandwidth centered at 1550~nm, on the CRed3 detector.
\end{itemize}

\section{Use of thermo-optic phase shifters for geometric phase compensation}
\label{sec:results}

\begin{figure}[h]
    \centering
    \includegraphics[width=0.6\textwidth]{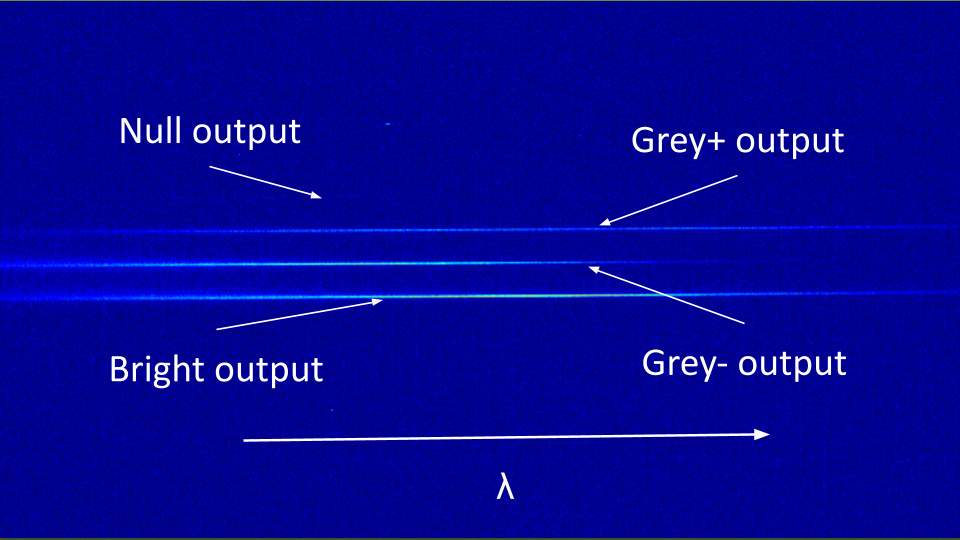}
    \caption{
    Spectrally dispersed output traces from the $4\times4$ MMI operating in the 2T-ABCD mode, demonstrating simultaneous capture of the phase quadrature states. The 2T-ABCD is configured on the white-light fringes with a full illumination of the bright output, a full extinction in the dark one and partial illumination on the grey outputs.}
     \label{fig:output_traces}
\end{figure}

Initial testing focused on evaluating the chromatic behavior of the TOPS and involved a $4\times4$ MMI operating in a 2-telescope injection configuration under active on-chip phase control.
In this operational mode, two distinct coherent beams are injected into the input ports of the MMI.
The geometry of the coupler generates four distinct interferometric outputs shifted in relative phase by increments of $90^\circ$, corresponding to a phase quadrature configuration, also called ABCD\cite{Lawson2000}.
The four outputs are classified by their interference states as the Null output, the Grey$+$ output, the Grey$-$ output and the Bright output (Fig.~\ref{fig:output_traces}).

\begin{figure}[h]
    \centering
    \includegraphics[width=0.6\textwidth]{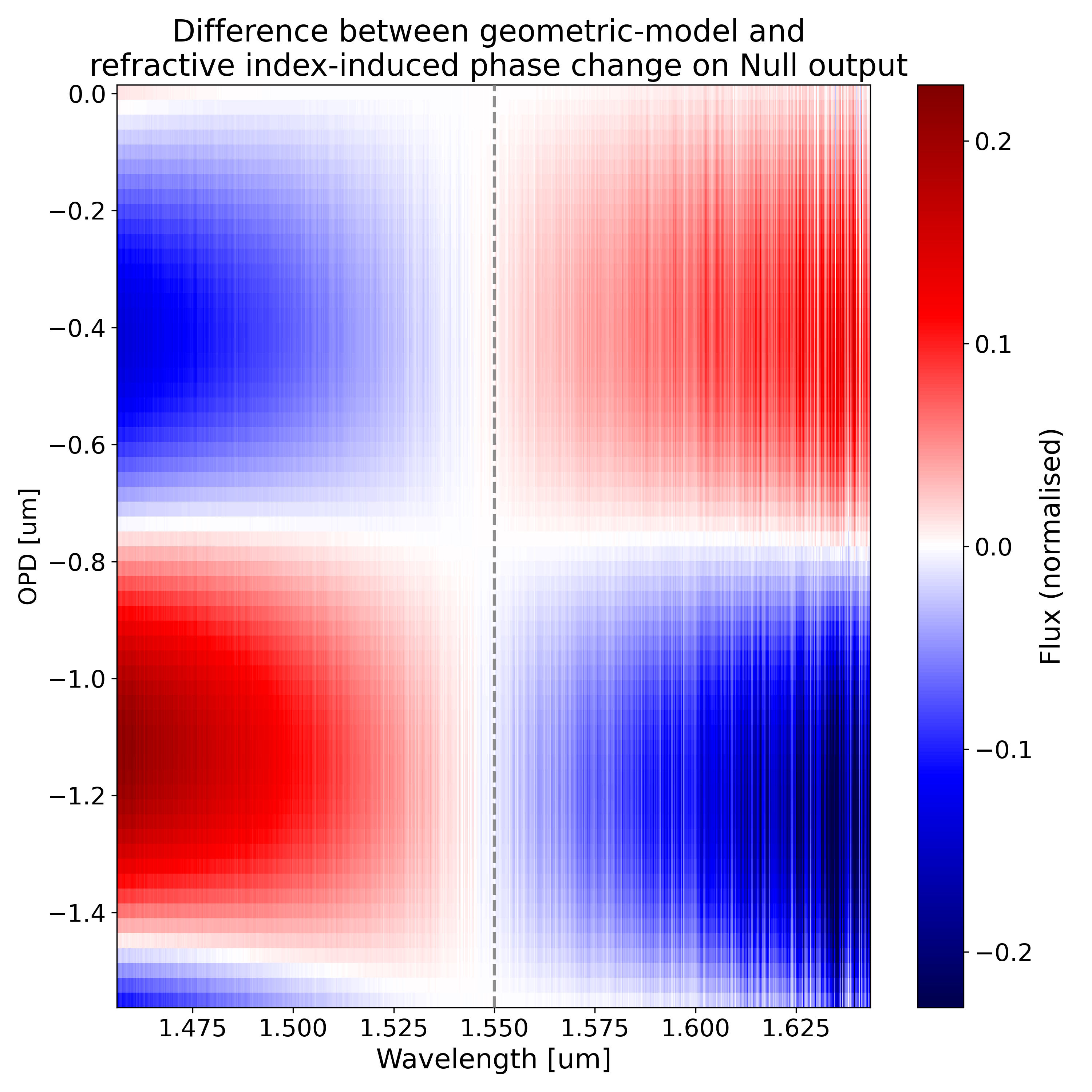}
    \caption{Measured phase deviation between the geometric path delay model and the actual thermo-optically induced phase change on the Null output waveguide as a function of wavelength.}
    \label{fig:dm_topa}
\end{figure}

A critical performance metric for astronomical nulling is the stability of the phase difference to $\pi$.
There is a dual challenge to reach this phase difference across all wavelengths and to maintain it.
Delay lines are usually used to get this phase and compensate for differential piston $\Delta \phi$ from atmospheric turbulence with fringe tracking.
However, such systems induce a geometric chromatic phase difference given by:
\begin{equation}
    \Delta \phi = 2 \pi \sigma \delta,
\end{equation}
with $\sigma$ the wavenumber and $\delta$ the Optical Path Difference (OPD) between the two beams.
The thermo-optic phase shifters could help to enhance the phase control with fine tuning within the chip.
To quantify the limitation imposed by the chromaticity of the optical effect, we mapped the phase difference between a purely geometric model and the actual refractive-index-induced phase change at the Null output across an operating band from $\lambda = 1.475\ \mu\text{m}$ to $1.625\ \mu\text{m}$ (Fig.~\ref{fig:dm_topa}).

The resulting error map indicates that the active on-chip thermo-optical modulators can efficiently track and counteract a geometric piston error across a localized spectral window of 50~nm.
Beyond this bandwidth, the differing chromatic characteristics of the glass substrate's thermo-optic coefficient and the geometric path delay diverge significantly, introducing structured phase leakage.

\section{Comparing null depths with 2 or 4 telescopes}
\label{sec:kernel}
\begin{figure}[h]
    \centering
    \includegraphics[width=0.7\linewidth]{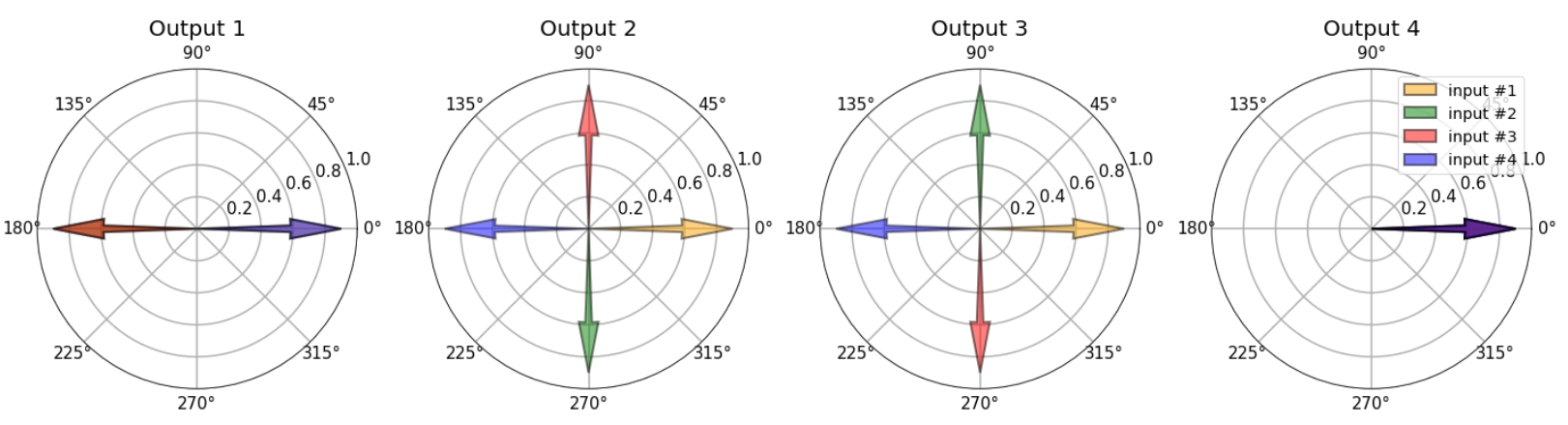}
    \caption{Theoretical complex matrix plot of the outputs of a $4 \times 4$ MMI. Input 1 is the phase reference for the other ones.\cite{Chingaipe2022}}
    \label{fig:mmi_phase}
\end{figure}

The combination of 4 telescopes in a $4 \times 4$ MMI (or in a staged-nuller like double-Bracewell architecture) delivers four different combinations of the inputs as shown on Figure~\ref{fig:mmi_phase}.
Output 1 combines the four beams to create two superimposed classic Bracewell null depths, called ``double'' null depth.
Outputs 2 and 3 form a conjugated pair that enables the creation of a kernel.
All beams constructively interfere in Output 4.

A systematic scan on three delay lines is performed to map the kernel space and find this position.
At least one pair of beams cannot reach the white fringe because of the internal OPD of its scanned beam.
A common range of piston values is defined so that the white fringes are encountered during the scan.
The MMI is then fed with the four beams.
The parameter space defined by a three-dimensional cube of piston ranges is explored.
For each triplet of DM pistons, the spectra of the four outputs are recorded.
The dark fringe is located in the datacube of the first output (Fig.~\ref{fig:output_traces}), over a restricted spectral bandwidth.
Consequently, the nulled signal is displayed in the null output. 

\begin{figure}[h]
    \centering
    \includegraphics[width=0.8\linewidth]{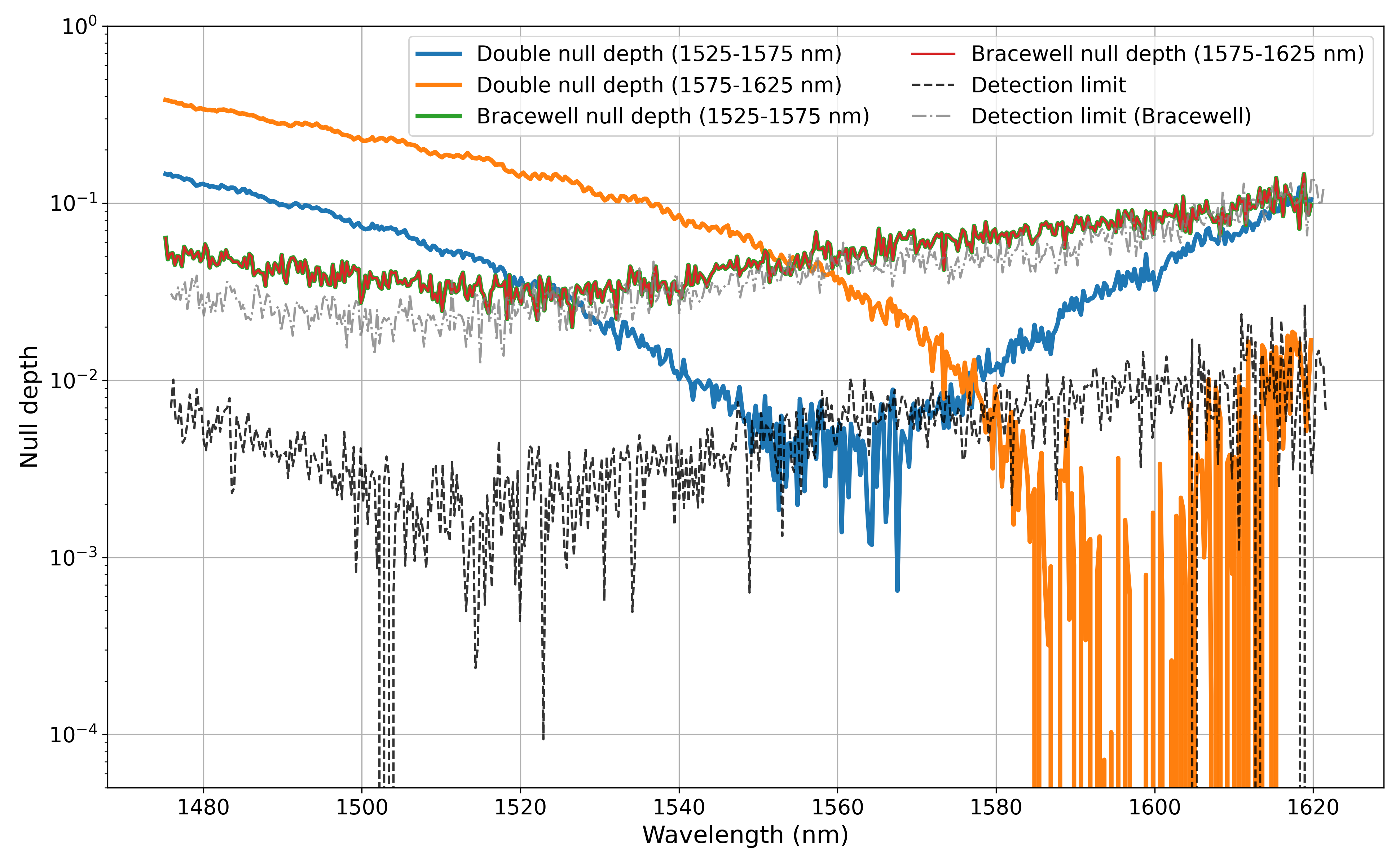}
    \caption{Spectra of null depths obtained with 4 and 2 telescopes. The dashed lines are the detection limits due to the detector noise.}
    \label{fig:2t_4t_comparison}
\end{figure}

The scans have been made in two configurations : with four beams and with two beams only.
In the first case, the null output encodes a ``double'' null (Fig.~\ref{fig:mmi_phase}, Output 1).
In the second, it encodes a classic Bracewell null depth.
As predicted\cite{Angel1997}, the nulled signal is deeper when it is the combination of two already nulled signals (Fig.~\ref{fig:2t_4t_comparison}).
The advantage of the MMI over a double-Bracewell architecture is the double null is obtained with a single device while the other architecture requires two stages with two and one coupler, respectively.
Moreover, Figure~\ref{fig:2t_4t_comparison} also shows the first case of a tunable nuller where it is possible to optimize the extinction at a particular wavelength.
This is of special interest where a deep contrast is needed to image a system at a particular line while still gathering information in the spectral continuum.
This figure also shows that the contrast limits are currently defined by the detector noise thus the absolute null depth reached on the bench is not a relevant value.

\section{Conclusion}
\label{sec:conclusions}
We have successfully implemented and evaluated the first laboratory prototypes of an integrated four-beam astronomical kernel-nuller based around a single monolithic MultiMode Interferometer coupler.
By pairing the natural alignment stability and compactness of a PIC chip with integrated, high-speed thermo-optical phase actuators, we achieved stable multi-port beam combination in a 2T-ABCD phase quadrature mode with no moving parts.
Our laboratory characterization shows that the thermo-optical effect can efficiently compensate for geometric OPD over a $50\text{~nm}$ bandwidth before their respective chromatic behaviors become significant.
The ``double'' null depth has been experimentally compared with a Bracewell one, for the first time.
The former exhibits a deeper extinction and allows fine tuning of nulling at a specific wavelength.

\section{Acknowledgments }
All the authors wish to acknowledge funding from the project PHOTONICS financed by the ANR program PEPR Origins (ANR-22-EXOR-0005).\\
This work was supported by the Action Spécifique Haute Résolution Angulaire (ASHRA) of CNRS/INSU co-funded by CNES.

\bibliography{report} 

@ARTICLE{Martinache2018,
       author = {{Martinache}, Frantz and {Ireland}, Michael J.},
        title = "{Kernel-nulling for a robust direct interferometric detection of extrasolar planets}",
      journal = {\aap},
         year = 2018,
        month = nov,
       volume = {619},
          eid = {A87},
        pages = {A87},
          doi = {10.1051/0004-6361/201832847},
archivePrefix = {arXiv},
       eprint = {1802.06252},
 primaryClass = {astro-ph.IM},
       adsurl = {https://ui.adsabs.harvard.edu/abs/2018A&A...619A..87M}
}

@INPROCEEDINGS{Chingaipe2022,
       author = {{Chingaipe}, Peter Marley and {Martinache}, Frantz and {Cvetojevic}, Nick},
        title = "{Four-input photonic kernel-nulling for the VLTI}",
    booktitle = {Optical and Infrared Interferometry and Imaging VIII},
         year = 2022,
       editor = {{M{\'e}rand}, Antoine and {Sallum}, Stephanie and {Sanchez-Bermudez}, Joel},
       series = {Society of Photo-Optical Instrumentation Engineers (SPIE) Conference Series},
       volume = {12183},
        month = aug,
          eid = {1218319},
        pages = {1218319},
          doi = {10.1117/12.2630050},
       adsurl = {https://ui.adsabs.harvard.edu/abs/2022SPIE12183E..19C}
}

@ARTICLE{Parra2024,
       author = {{Parra}, Jorge and {Navarro-Arenas}, Juan and {Sanchis}, Pablo},
        title = "{Silicon thermo-optic phase shifters: a review of configurations and optimization strategies}",
      journal = {Advanced Photonics Nexus},
         year = 2024,
        month = jul,
       volume = {3},
          eid = {044001},
        pages = {044001},
          doi = {10.1117/1.APN.3.4.044001},
       adsurl = {https://ui.adsabs.harvard.edu/abs/2024AdPhN...3d4001P}
}

@PROCEEDINGS{Lawson2000,
        title = "{Principles of Long Baseline Stellar Interferometry}",
    booktitle = {Principles of Long Baseline Stellar Interferometry},
         year = 2000,
       editor = {{Lawson}, Peter R.},
        month = jan,
       adsurl = {https://ui.adsabs.harvard.edu/abs/2000plbs.conf.....L}
}

@ARTICLE{martinod2018,
       author = {{Martinod}, M.~A. and {Mourard}, D. and {B{\'e}rio}, P. and {Perraut}, K. and {Meilland}, A. and {Bailet}, C. and {Bresson}, Y. and {ten Brummelaar}, T. and {Clausse}, J.~M. and {Dejonghe}, J. and {Ireland}, M. and {Millour}, F. and {Monnier}, J.~D. and {Sturmann}, J. and {Sturmann}, L. and {Tallon}, M.},
        title = "{Fibered visible interferometry and adaptive optics: FRIEND at CHARA}",
      journal = {Astron. Astrophys},
         year = 2018,
        month = oct,
       volume = {618},
          eid = {A153},
        pages = {A153},
          doi = {10.1051/0004-6361/201731386},
       adsurl = {https://ui.adsabs.harvard.edu/abs/2018A&A...618A.153M}
}

@ARTICLE{huby2012,
       author = {{Huby}, E. and {Perrin}, G. and {Marchis}, F. and {Lacour}, S. and {Kotani}, T. and {Duch{\^e}ne}, G. and {Choquet}, E. and {Gates}, E.~L. and {Woillez}, J.~M. and {Lai}, O. and {F{\'e}dou}, P. and {Collin}, C. and {Chapron}, F. and {Arslanyan}, V. and {Burns}, K.~J.},
        title = "{FIRST, a fibered aperture masking instrument. I. First on-sky test results}",
      journal = {Astron. Astrophys},
         year = 2012,
        month = may,
       volume = {541},
          eid = {A55},
        pages = {A55},
          doi = {10.1051/0004-6361/201118517},
archivePrefix = {arXiv},
       eprint = {1203.5075},
 primaryClass = {astro-ph.IM},
       adsurl = {https://ui.adsabs.harvard.edu/abs/2012A&A...541A..55H}
}

@ARTICLE{gravity2017,
       author = {{Gravity Collaboration}},
        title = "{First light for GRAVITY: Phase referencing optical interferometry for the Very Large Telescope Interferometer}",
      journal = {Astron. Astrophys},
         year = 2017,
        month = jun,
       volume = {602},
          eid = {A94},
        pages = {A94},
          doi = {10.1051/0004-6361/201730838},
archivePrefix = {arXiv},
       eprint = {1705.02345},
 primaryClass = {astro-ph.IM},
       adsurl = {https://ui.adsabs.harvard.edu/abs/2017A&A...602A..94G}
}

@ARTICLE{Bracewell1978,
       author = {{Bracewell}, R.~N.},
        title = "{Detecting nonsolar planets by spinning infrared interferometer}",
      journal = {Nature},
         year = 1978,
        month = aug,
       volume = {274},
       number = {5673},
        pages = {780-781},
          doi = {10.1038/274780a0},
       adsurl = {https://ui.adsabs.harvard.edu/abs/1978Natur.274..780B}
}

@ARTICLE{martinod2021,
       author = {{Martinod}, M-.A. and {Norris}, B. and {Tuthill}, P. and {Lagadec}, T. and {Jovanovic}, N. and {Cvetojevic}, N. and Gross, S. and {Arriola}, A. and {Gretzinger}, T. and {Withford}, M. J. and {Guyon}, O. and {Lozi}, J. and {Lawrence}, J. S. and {Leon-Saval}, S.},
        title = "{Scalable photonic-based nulling interferometry with the dispersed, multi-baseline GLINT instrument}",
      journal = {Nature Communications},
         year = 2021,
        month = apr,
       volume = {12},
        pages = {2465},
          doi = {10.1038/s41467-021-22769-x}
}

@ARTICLE{2019A&A...623L..11G,
       author = {{Gravity Collaboration}},
        title = "{First direct detection of an exoplanet by optical interferometry. Astrometry and K-band spectroscopy of HR 8799 e}",
      journal = {Astron. Astrophys},
         year = 2019,
        month = mar,
       volume = {623},
          eid = {L11},
        pages = {L11},
          doi = {10.1051/0004-6361/201935253},
archivePrefix = {arXiv},
       eprint = {1903.11903},
 primaryClass = {astro-ph.EP},
       adsurl = {https://ui.adsabs.harvard.edu/abs/2019A&A...623L..11G}
}

@ARTICLE{Marois2008,
       author = {{Marois}, Christian and {Macintosh}, Bruce and {Barman}, Travis and
         {Zuckerman}, B. and {Song}, Inseok and {Patience}, Jennifer and
         {Lafreni{\`e}re}, David and {Doyon}, Ren{\'e}},
        title = "{Direct Imaging of Multiple Planets Orbiting the Star HR 8799}",
      journal = {Science},
         year = 2008,
        month = nov,
       volume = {322},
       number = {5906},
        pages = {1348},
          doi = {10.1126/science.1166585},
archivePrefix = {arXiv},
       eprint = {0811.2606},
 primaryClass = {astro-ph},
       adsurl = {https://ui.adsabs.harvard.edu/abs/2008Sci...322.1348M}
}

@ARTICLE{Schworer2015,
       author = {{Schworer}, Guillaume and {Tuthill}, Peter G.},
        title = "{Predicting exoplanet observability in time, contrast, separation, and polarization, in scattered light}",
      journal = {Astron. Astrophys},
         year = 2015,
        month = jun,
       volume = {578},
          eid = {A59},
        pages = {A59},
          doi = {10.1051/0004-6361/201424202},
archivePrefix = {arXiv},
       eprint = {1505.03082},
 primaryClass = {astro-ph.IM},
       adsurl = {https://ui.adsabs.harvard.edu/abs/2015A&A...578A..59S}
}

@INPROCEEDINGS{Spalding2024,
       author = {{Spalding}, Eckhart and {Arcadi}, Elizabeth and {Douglass}, Glen and {Gross}, Simon and {Guyon}, Olivier and {Martinod}, Marc-Antoine and {Norris}, Barnaby and {Rossini-Bryson}, Stephanie and {Taras}, Adam and {Tuthill}, Peter and {Ahn}, Kyohoon and {Deo}, Vincent and {El Morsy}, Mona and {Lozi}, Julien and {Vievard}, Sebastien and {Withford}, Michael},
        title = "{The GLINT nulling interferometer: improving nulls for high-contrast imaging}",
    booktitle = {Optical and Infrared Interferometry and Imaging IX},
         year = 2024,
       editor = {{Kammerer}, Jens and {Sallum}, Stephanie and {Sanchez-Bermudez}, Joel},
       series = {Society of Photo-Optical Instrumentation Engineers (SPIE) Conference Series},
       volume = {13095},
        month = aug,
          eid = {1309507},
        pages = {1309507},
          doi = {10.1117/12.3016348},
       adsurl = {https://ui.adsabs.harvard.edu/abs/2024SPIE13095E..07S}
}

@INPROCEEDINGS{Rossini2026,
       author = {{Rossini-Bryson}, Stephanie and {Norris}, B. and {Tuthill}, P. and {Arcadi}, E. and {Lozi}, J. and {Guyon}, O. and {Gross}, S.},
        title = "{The GLINT instrument: high-contrast imaging using nulling interferometry with photonic chips}",
    booktitle = {Planetary formation and Exoplanets in the ELT era (Exo-ELT)},
         year = 2026,
        month = apr,
          eid = {51},
        pages = {51},
          doi = {10.5281/zenodo.19686477},
       adsurl = {https://ui.adsabs.harvard.edu/abs/2026exoe.confE..51R}
}

@INPROCEEDINGS{hi5_defrere,
       author = {{Defr\`ere}, D. and {Absil}, O. and {Berger}, J.-P. and {Bigioli}, A. and {Courtney-Barrer}, B. and {Dandumont}, C. and {Emsenhuber}, A. and {Ertel}, S. and {Gagne}, J. and {Garreau}, G. and {Glauser}, A. and {Gross}, S. and {Ireland}, M. and {Kenchington}, H.-D. and {Kraus}, S. and {Labadie}, L. and {Laborde}, V. and {Laugier}, R. and {Leisenring}, J. and {Loicq}, J. and {Martin}, G. and {Martinache}, F. and {Martinod}, M.-A. and {Matter}, A. and {Mazzoli}, A. and {Mennesson}, B. and {Salman}, M. and {Raskin}, G. and {Vandenbussche}, B. and {Verlinden}, S. and {Woillez}, J.},
        title = "{L-band nulling interferometry at the VLTI with ASGARD/Hi-5: status and plans}",
    booktitle = {Optical and Infrared Interferometry and Imaging VIII},
         year = 2022,
       series = {Society of Photo-Optical Instrumentation Engineers (SPIE) Conference Series},
       volume = {12183},
        month = jul,
          eid = {12183-16},
        pages = {12183-16}
}

@INPROCEEDINGS{nott_sanny2026_perf_chip,
       author = {{Sanny}, A. and {Labadie}, L. and {Gross}, S. and {Barjot}, K. and {Garreau}, G. and {Laugier}, R. and {Martinod}, M.-A. and {Mattheussen}, T. and {Chingaipe}, P. and {Withford}, M. and {Defr\`ere}, D.},
        title = "{Asgard/NOTT: the performance of the integrated optics 4-telescope beam combiner for double-Bracewell nulling interferometry}",
    booktitle = {Optical and Infrared Interferometry and Imaging X},
         year = 2026,
       series = {Society of Photo-Optical Instrumentation Engineers (SPIE) Conference Series},
       volume = {14148},
        month = jul,
          eid = {14148-8},
        pages = {14148-8}
}

@INPROCEEDINGS{dahl_seidr2026,
       author = {{Dahl}, D. S. and {Long}, N. K. and {Betters}, C. H. and {Bryant}, J. J. and {Cvetojevic}, N. and {Ireland}, M. J. and {Kraus}, S. and {Leon-Saval}, S. and {Martinache}, F. and {Martinod}, M.-A. and {Norris}, B. and {Paul}, J. and {Rodziewicz-Ryan}, A. and {Taras}, A. K. and {Wei}, J. and {Tuthill}, P. G.},
        title = "{Seidr update: photonic ‘black magic’ for high-contrast interferometry using kernel-nulling and photonic lanterns}",
    booktitle = {Optical and Infrared Interferometry and Imaging X},
         year = 2026,
       series = {Society of Photo-Optical Instrumentation Engineers (SPIE) Conference Series},
       volume = {14148},
        month = jul,
          eid = {14148-82},
        pages = {14148-82}
}

@ARTICLE{Cvetojevic2022,
       author = {{Cvetojevic}, Nick and {Martinache}, Frantz and {Chingaipe}, Peter and {Laugier}, Romain and {{\L}awniczuk}, Katarzyna and {Broeke}, Ronald G. and {Ligi}, Roxanne and {N'Diaye}, Mamadou and {Mary}, David},
        title = "{3-beam self-calibrated Kernel nulling photonic interferometer}",
      journal = {arXiv e-prints},
         year = 2022,
        month = jun,
          eid = {arXiv:2206.04977},
        pages = {arXiv:2206.04977},
          doi = {10.48550/arXiv.2206.04977},
archivePrefix = {arXiv},
       eprint = {2206.04977},
 primaryClass = {astro-ph.IM},
       adsurl = {https://ui.adsabs.harvard.edu/abs/2022arXiv220604977C}
}

@ARTICLE{Angel1997,
       author = {{Angel}, J.~R.~P. and {Woolf}, N.~J.},
        title = "{An Imaging Nulling Interferometer to Study Extrasolar Planets}",
      journal = {\apj},
         year = 1997,
        month = jan,
       volume = {475},
       number = {1},
        pages = {373-379},
          doi = {10.1086/303529},
       adsurl = {https://ui.adsabs.harvard.edu/abs/1997ApJ...475..373A}
}

@ARTICLE{norris2020,
       author = {{Norris}, Barnaby R.~M. and {Cvetojevic}, Nick and {Lagadec}, Tiphaine and
         {Jovanovic}, Nemanja and {Gross}, Simon and {Arriola}, Alexander and
         {Gretzinger}, Thomas and {Martinod}, Marc-Antoine and {Guyon}, Olivier and
         {Lozi}, Julien and {Withford}, Michael J. and {Lawrence}, Jon S. and
         {Tuthill}, Peter},
        title = "{First on-sky demonstration of an integrated-photonic nulling interferometer: the GLINT instrument}",
      journal = {\mnras},
         year = 2020,
        month = jan,
       volume = {491},
       number = {3},
        pages = {4180-4193},
          doi = {10.1093/mnras/stz3277},
archivePrefix = {arXiv},
       eprint = {1911.09808},
 primaryClass = {astro-ph.IM},
       adsurl = {https://ui.adsabs.harvard.edu/abs/2020MNRAS.491.4180N}
}

@INPROCEEDINGS{lagadec2018,
       author = {{Lagadec}, Tiphaine and {Norris}, Barnaby and {Gross}, Simon and
         {Arriola}, Alexander and {Gretzinger}, Thomas and {Cvetojevic}, Nick and
         {Lawrence}, Jon and {Withford}, Michael and {Tuthill}, Peter},
        title = "{GLINT South: a photonic nulling interferometer pathfinder at the Anglo-Australian Telescope for high contrast imaging of substellar companions}",
    booktitle = {\procspie},
         year = 2018,
       series = {Society of Photo-Optical Instrumentation Engineers (SPIE) Conference Series},
       volume = {10701},
        month = jul,
          eid = {107010V},
        pages = {107010V},
          doi = {10.1117/12.2313171},
       adsurl = {https://ui.adsabs.harvard.edu/abs/2018SPIE10701E..0VL}
}
\bibliographystyle{spiebib} 

\end{document}